\documentclass[aps,prl,twocolumn,superscriptaddress]{revtex4-2}
\usepackage{bm}                 
\usepackage{graphicx}
\usepackage{xcolor}
\usepackage{float}
\newcommand{\Chula}{Department of Physics, Faculty of Science, Chulalongkorn University, Patumwan, Bangkok 10330, Thailand}

\begin{document}
\title{Magnetoconductance Oscillations in Electron-hole Hybridization Gaps and Valley Splittings in Tetralayer Graphene}

\author{Illias Klanurak}
\affiliation{\Chula}

\author{Kenji Watanabe}
\affiliation{Research Center for Functional Materials, National Institute for Materials Science, 1-1 Namiki, Tsukuba 305-0044, Japan.}

\author{Takashi Taniguchi}
\affiliation{International Center for Materials Nanoarchitectonics, National Institute for Materials Science, 1-1 Namiki, Tsukuba 305-0044, Japan.}

\author{Sojiphong Chatraphorn}
\affiliation{\Chula}

\author{Thiti Taychatanapat}
\email[]{thiti.t@chula.ac.th}
\affiliation{\Chula}

\date{\today}
\begin{abstract}
	We investigate  magnetotransport on Bernal-stacked tetralayer graphene whose band structure consists of two massive subbands with different effective masses. Under a finite displacement field, we observe valley splitting of Landau levels (LLs) only in the light-mass subband, consistent with a tight-binding model. At low density, we find unexpected magnetoconductance oscillations in bulk gaps which originate from a series of hybridizations between electron-like and hole-like LLs due to band inversion in tetralayer graphene. In contrast to a trivial LL quantization gap, these inverted hybridization gaps can lead to a change in number of edge states which explains the observed oscillations.
\end{abstract}

\maketitle

The electronic properties of Bernal-stacked tetralayer graphene (4LG) have been shown to exhibit many intriguing phenomena such as insulating state~\cite{Grushina2015insulating4LG}, interlayer interaction~\cite{Wu2015ABABinterlayer}, tunable Lifshitz transitions~\cite{shi2018tunable}, helical edge states~\cite{che2020helical}, and unconventional satellite peaks~\cite{Mukai2021UnconventionalABAB}. Its band structure features band inversion as a result of band overlap and hybridization between the two bilayer-graphene-like (BLG-like) subbands~\cite{Latil2006Bandstruc,Aoki2007Bandstruc,koshino2011landau}. Bernal-stacked trilayer graphene (TLG) also possesses a band overlap but it occurs at a narrow density range which is hard to resolve experimentally. In addition, the subbands in TLG only hybridize  in a presence of perpendicular electric field, while those in 4LG always hybridize due to the next-nearest interlayer hopping parameters~\cite{koshino2011landau}.

The band inversion in 4LG provides a unique platform to study quantum Hall effect (QHE) in a regime in which both electron-like and hole-like states coexist. Previously, the energy inversion of electron-like and hole-like Landau levels has been engineered by electric field or both in-plane and out-of-plane magnetic fields to form helical edge states~\cite{young2014tunable,sanchez2017helical,che2020helical}. Some of two-dimensional quantum wells also host inverted electron-hole system which, under a magnetic field, reveals novel aspects of QHE~\cite{Suzuki2004LLhybridization,Gusev2010electronHole} and quantum spin Hall effect~\cite{Konig2007,Du2015HelicalEdge,Ma2015UnexpectedEdge,Karalic2016InvertedLL, Yahniuk2019Inverted,Chen2019Interaction}.

\begin{figure}
	\includegraphics{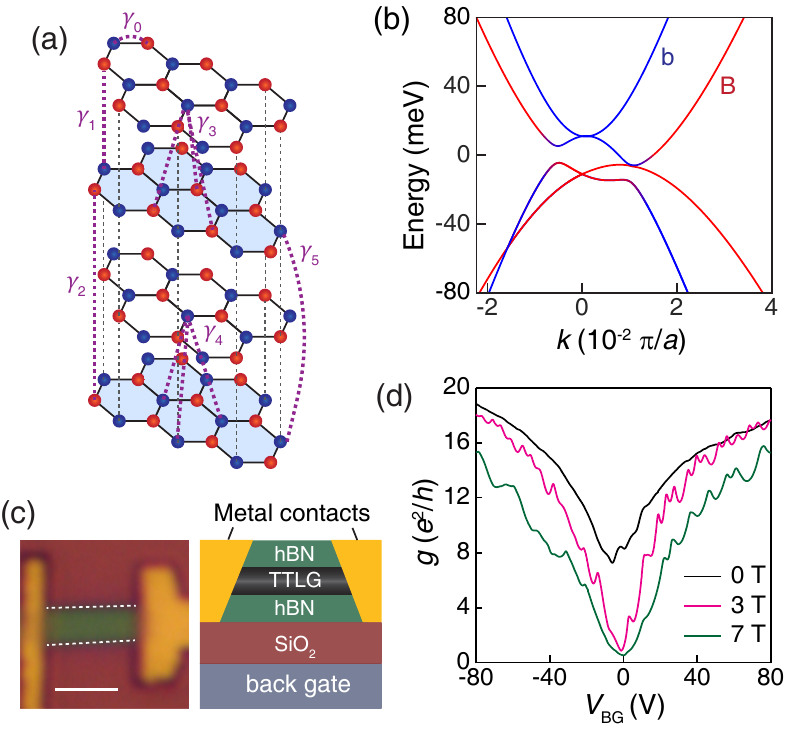}
	\caption{(a) Crystal structure of 4LG. The purple dashed lines indicate two hopping sites for $\gamma_i$ parameters.  (b) Band structure of 4LG from TB model. The  heavy- and light-mass bands are labeled as B (red) and b (blue), respectively. The lattice constant $a$ is $0.246$~nm. (c) Left: an optical image of an encapsulated 4LG device. Scale bar, $2$~$\mu$m. Right: schematic diagram illustrating a cross-sectional view of the 4LG device. (d) Two-terminal conductance as a function of $V_{\rm BG}$ at $0$ (black), $3$ (pink), and $7$~T (green), respectively. \label{fig:bandstruc}}
	
\end{figure}


In this work, we investigate magnetotransport properties of 4LG encapsulated by hexagonal boron nitride (hBN). The electronic band structure of 4LG can be characterized by a set of hopping parameters $\gamma_{0}$-$\gamma_{5}$ [Fig.~$1$(a)] and energy imbalance between dimmer and non-dimmer sites $\delta$~\cite{koshino2011landau}. The low energy band structure of 4LG, shown in Fig.~$1$(b), comprises two BLG-like subbands with different effective masses~\cite{koshino2010interlayer,koshino2011landau,yagi2018low}. We denote the light-mass subband by b and the heavy-mass subband by B.  The presence of skewed lattice sites hopping $\gamma_{3}$ induces trigonal warping. The next-nearest interlayer hopping parameters $\gamma_{2}$ and $\gamma_{5}$ cause the b and B subbands to overlap and hybridize at low energy as shown in Fig.~$1$(b)~\cite{koshino2011landau, shi2018tunable}. The remaining hopping parameters $\gamma_{4}$ and $\delta$ generate electron-hole asymmetry in the band structure.

\begin{figure*}
	\includegraphics{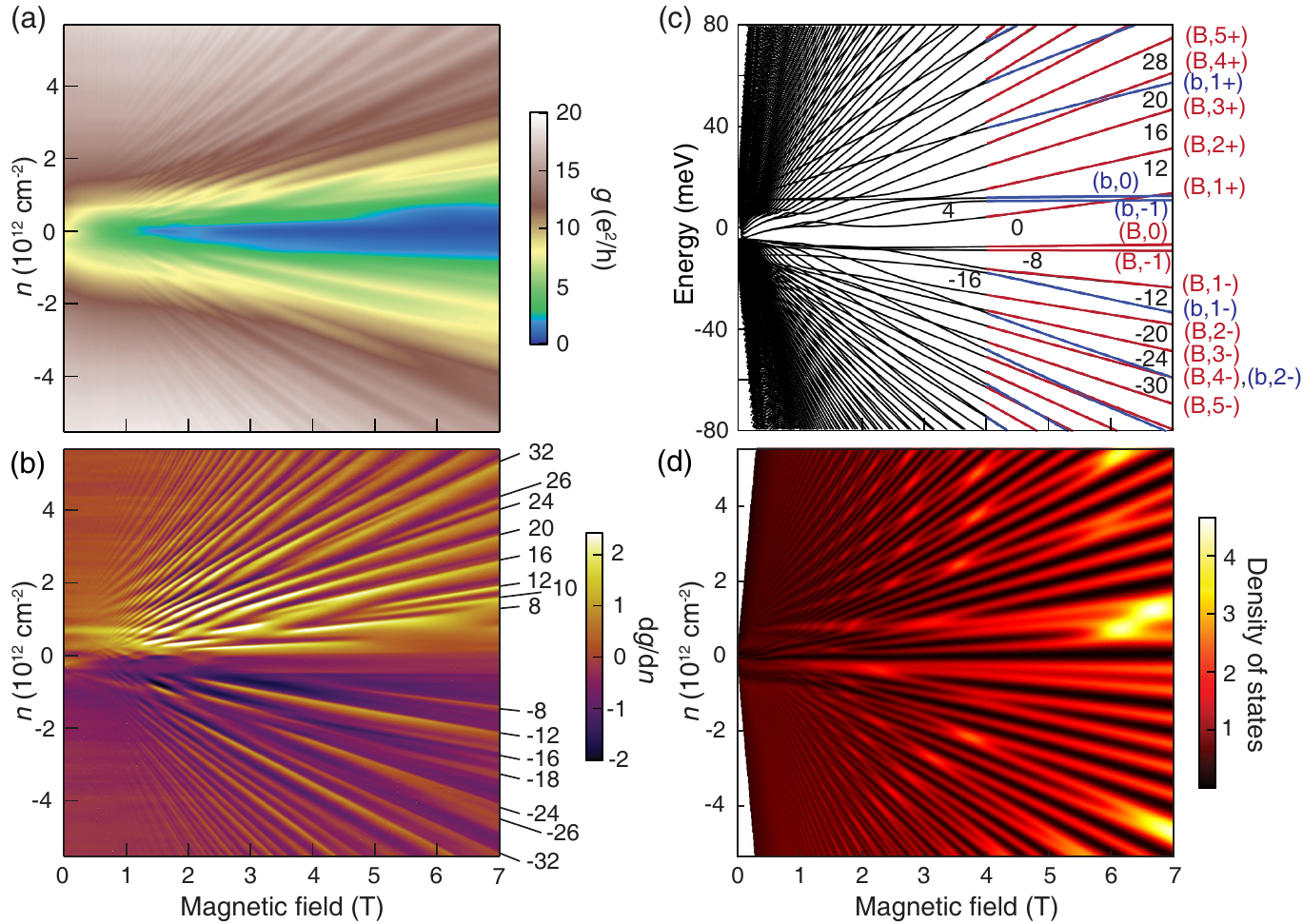}
	\caption{(a) conductance as a function of $n$ and $B$. (b) ${\rm d} g/ {\rm d} n$ numerically calculated from data in (a). The numbers at the right border indicate filling factors. (c) LL spectra in 4LG calculated from TB model. The filling factors associated with some energy gaps are labeled. The labels at the right border denote quantum states for each LL. At high field, LLs from the b and B subbands are in blue and red, respectively. At low field, LLs from both subbands are in black due to LL hybridization. (d) Calculated density of states of 4LG as a function of $n$ and $B$. \label{fig:wholeLL}}
\end{figure*}

In our devices, we use a dry transfer technique to assemble the heterostructures and etch them in CHF$_3$/O$_2$ plasma. Edge contacts are defined by electron beam lithography and formed by sputtering 80-nm Mo~\cite{wang2013one}.  The standard lock-in technique is used to investigate magnetotransport properties. All measurements are performed at 2.4~K unless stated otherwise. Figure~$1$(c) shows an optical image and a schematic diagram of a device.


We first examine two-terminal conductance $g$ as a function of back-gate voltage $V_{\rm BG}$ at magnetic field $B=0$ [Fig.~$1$(d), black line]. We find that the conductance curve exhibits multiple local minimums, which are associated with band edges and Lifshitz transitions near zero energy in the band structure of 4LG as shown in Fig.~$1$(b)~ \cite{shi2018tunable,Hirahara20184LGpeaks}. At finite $B$, we observe conductance oscillations in $V_{\rm BG}$ due to the QHE [Fig.~$1$(d)]. Our conductance does not develop into well-defined plateaus even at $7$~T, likely due to relatively high contact resistance ($\sim$1~k$\Omega$) and geometry effect on two-terminal conductance which leads to the distortion of the QH plateaus \cite{abanin2008conformal,williams2009quantum}. At high field, the conductance exhibits a single minimum at which we associate to the charge neutrality point (CNP) of the sample. Near-zero gate voltage for the CNP indicates pristine quality of our samples.

To further investigate the magnetotransport properties of the system, we measure $g$ as a function of $V_{\rm BG}$ and $B$ [Fig.~$2$(a)]. Charge carrier density $n$ is determined from the period of conductance oscillations at high $B$. The Landau fan diagram displays rich features associated with LLs from the b and B subbands. To see the oscillations more clearly, we calculate ${\rm d} g/ {\rm d} n$ [Fig.~$2$(b)]. The dark lines correspond to energy gaps with associated filling factors $\nu$ shown in the figure. Multiple LL crossings are evident. For instance, the dark line for $\nu=12$, seen clearly at $7$~T, disappears at about $5$~T and reemerges again at $3$~T. The absence of the dark line for $\nu = 12$ between $3$ and $5$ T is the result of the LL crossing.

To gain a better understanding of the Landau fan diagram, we calculate energies of LLs as a function of $B$ using the tight-binding (TB) model [Fig.~$2$(c)]. In this plot, we assume that potential difference between layers is zero and use $\gamma_{0}=3.1$, $\gamma_{1}=0.39$, $\gamma_{2}=-0.022$, $\gamma_{3}=0.315$, $\gamma_{4}=0.12$, $\gamma_{5}=0.018$ and $\delta=0.020$ eV. These TB parameters are determined by matching the LL crossing positions from the experiment at low density with those from the calculation~\cite{taychatanapat2011quantum}. We note that our device has a single gate. As we induce higher carrier density via back gate, 4LG is inevitably subject to a stronger displacement field which induces larger potential difference between layers. Therefore, our simulation gives a good agreement with data from low density at which potential difference is still small.

From the spectra, LLs of 4LG at high $B$ can be viewed as a combination of two sets of BLG-like LLs from the b and B subbands~\cite{koshino2011landau,yin2017landau}. At high energy, a LL energy is approximately linear in $B$, as expected from bilayer nature of the subbands. The energy spacing of the LLs from the light-mass band b is larger than that of the heavy-mass band B because cyclotron frequency is inversely proportional to effective mass. The mixing between LLs due to $\gamma_2$ and $\gamma_5$ parameters and trigonal warping effect from $\gamma_3$ parameter leads to hybridization gaps, more visible at low energy~\cite{koshino2011landau}. We label each LL in Fig.~$2$(c) with two indices, indicating the subband (B or b) and LL index $n$ ($n+$ for electron-like and $n-$ for hole-like LLs). For the zeroth index LL, we label them as (b/B, 0) and (b/B, $-1$).  These two LLs are degenerate in bilayer graphene but the degeneracy is slightly lifted in 4LG with a small energy gap of $\sim$2~meV at $7$~T. The numbers inside LL energy gaps in Fig.~$2$(c) indicate values of $\nu$ associated with the gaps. Each LL has four-fold spin and valley degeneracy~\cite{koshino2011landau}.

To compare our data with the calculation, we simulate density of states as a function of $B$ and $n$ from the LL spectra in Fig.~$2$(c) to obtain the plot in Fig.~$2$(d). Here,  we assume a Lorentzian line-shape for each LL with broadening of 1.5 meV, estimated from a LL gap from the b subband at the onset of the oscillations of $0.5$~T [Fig.~$2$(b)]. The simulation captures  main features of the experimental data in Fig.~$2$(b) such as the positions of LL crossing at low density and a position of the horizontal line in the electron side which originates from the zeroth LL of the light-mass band.

However, some discrepancies exist between our data and simulation. In our data, some LLs are two-fold degenerate. For example, in Fig.~$2$(b) at $6$~T, we observe the filling factor sequences of \{$8$, $10$, $12$\} and \{$24$, $26$, $32$\} which imply LLs with degeneracy of $2$ and $6$. Comparing the positions of these LLs with the LL spectra in Fig.~$2$(c), we conclude that each of the 4-fold degenerate (b,~$0$) and (b,~$1+$) split into $2$ LLs with $2$-fold degeneracy while the LLs from the B subband remain 4-fold degenerate.  The $6$-fold degeneracy observed when filling factors change from $26$ to $32$ is the result of LL crossing between the $2$-fold degenerate LL from the splitting of (b,~$1+$) and the $4$-fold degenerate (B,~$5+$).

\begin{figure}
	\centering
	\includegraphics{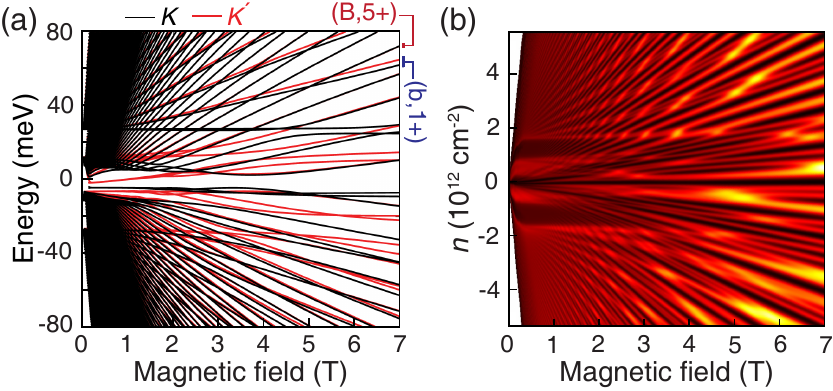}
	\caption{(a) LLs energy of the bulk 4LG as a function of $B$ with potential difference between adjacent layers of $16$~meV. The black and red lines are LLs for $K$ and $K'$, respectively. (b) Calculated density of states of the LLs in (a) as a function of $n$ and $B$.}
	\label{fig:finiteU}
\end{figure}

In 4LG, a 2-fold degenerate LL can occur by applying a displacement field to generate a potential difference between layers which breaks inversion symmetry and lifts valley degeneracy. To capture the effect of the potential difference on LLs, we simulate LL spectra  using a constant value of the potential difference between adjacent layers of  $16$~meV and the same set of the hopping parameters used in Fig.~$2$(c). We note that, in our measurement, a value of potential difference will vary as we change density (see more details in Supplemental Material~\cite{RefSM}). The black and red lines in Fig.~$3$(a) represent LLs from $K$ and $K'$ valleys, respectively. The valley splittings of the LLs from the B~subband are much smaller than LL broadening and therefore they continue to appear $4$-fold degenerate in our measurement. However, the splittings are much more pronounced for LLs from the b subband, consistent with the data in which  the LL splittings are observed in the b subband only. Although we can explain most observed features within a single-electron picture, we cannot completely rule out interaction-induced LL splitting. For example, the (b,~$0$) and (b,~$-1$)  are so close in energy that they should experience similar value of the potential difference. However, we observe the splitting of (b,~$0$) but not (b,~$ -1$) which may suggest that other symmetry-breaking mechanisms are involved.

\begin{figure*}
    \centering
    \includegraphics{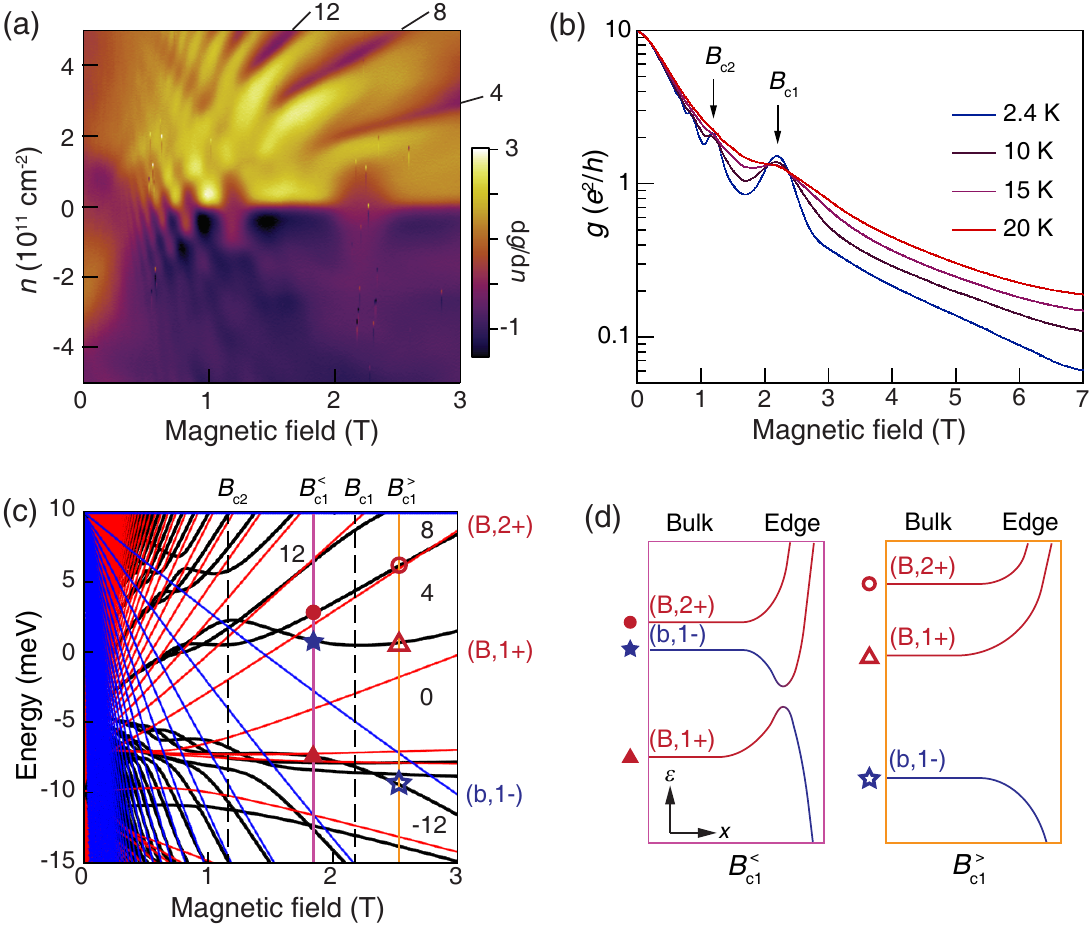}
    \caption{(a) Color map of ${\rm d} g/{\rm d} n$ at low $n$ and $B$. The integer numbers indicate filling factors of the dark diagonal lines. (b) Conductance oscillations at $n = 0$ as a function of $B$ at various temperatures. The temperature at which the oscillations disappear ($\sim$20~K) is consistent with the size of the $\nu = 0$ gap. (c) The bulk LL spectra of 4LG (black lines). The blue and red lines are LL spectra of b and B subbands, respectively, calculated without band hybridization. (d) Diagrams  of LL edge states at $B_{c1}^<$ and $B_{c1}^>$, indicated by pink and orange lines in (c). We omit the zeroth LLs (B,~$0$) and (B,~$-1$) (two red horizontal lines). }
    \label{fig:zeroFF}
\end{figure*}

A more surprising discrepancy between our data and the simulation occurs at zero filling factor. Figure~\ref{fig:zeroFF}(a) shows a plot of ${\rm d}g/{\rm d}n$ at low $n$ and $B$ while Figure~\ref{fig:zeroFF}(b) displays $g$ at $n = 0$ as a function of $B$ at various temperatures. We observe oscillations in magnetoconductance clearly along the zero density in all three devices we have measured (see data in Supplemental Material~\cite{RefSM}). Typically, the conductance oscillations in QHE occur when Fermi energy passes through a LL.  However, from the calculation of the LL spectra from the TB model [Fig.~\ref{fig:zeroFF}(c), black lines], there is no LL crossing inside $\nu = 0$. As a result, the conductance should exhibit no oscillation in $B$ at $\nu = 0$, contradicting our results.

To resolve this discrepancy, we examine the complex nature of low-energy LL spectra. Due to the band inversion in 4LG at zero magnetic field [see Fig.~\ref{fig:bandstruc}(b)],  hole states from the b subband reside at higher energy than electron states from the B subband for wave vectors around zero. As $B$ increases, energies of hole-like LLs decrease while those of electron-like LLs increase. The opposite magnetic dependence of LL energies leads to a series of crossings and anti-crossings which manifests as three adjacent LLs braided together at low field [see Fig.~4(c), black lines]. These anti-crossings, whose energy gaps depend on $\gamma_2$ and $\gamma_5$ parameters, are the result of hybridization between hole-like LLs and electron-like LLs from the b and B subbands, respectively. A braiding of three LLs is the consequence of the trigonal warping effect from $\gamma_3$ parameter.  The effect causes the anti-crossings to occur when the LL indices of the unperturbed LLs are the same or differ by multiples of $3$~\cite{Serbyn2013hybrid,campos2016TLG, Shimazaki2016LandauTrilayer,Zibrov2018Gully}.

To identify the underlying LLs that hybridize into the braided LLs, we calculate LL spectra by setting the mixing terms between the subbands to zero~\cite{koshino2011landau}. The result is displayed in Fig.~\ref{fig:zeroFF}(c). The blue and red lines represent LLs of the b and B subbands, respectively. Comparing the LLs with and without the mixing terms, we find that the energy gap at $\nu = 0$ originates from two different mechanisms with a crossover at the critical field $B_{c1}$ of $\sim$2.1~T. For $B > B_{c1}$, the $\nu=0$ gap is a trivial LL gap which is always present regardless of LL hybridization. For $B < B_{c1}$, the $\nu=0$ gap emerges from a series of hybridizations between hole-like (b,~$i-$) and electron-like (B,~$i+$) for $i = 1,2,\ldots$.

Let us focus on a $\nu = 0$ gap at $B_{c1}$ arising from the hybridization between (b,~$1-$) and (B,~$1+$). At $B_{c1}^< < B_{c1}$, the hybridized LLs at higher energy (solid blue star) and lower energy (solid red triangle) in Fig.~\ref{fig:zeroFF}(c) are mostly dominated by hole-like (b,~$1-$) and electron-like (B,~$1+$), respectively. As we increase $B$ beyond $B_{c1}$, the admixture of each hybridized LL gradually changes and the situation becomes reversed. Now, the higher-energy LL (hollow red triangle) evolves into the electron-like (B,~$1+$) while the lower-energy LL (hollow blue star) turns into the hole-like (b,~$1-$).  Therefore, as we increase $B$, the characteristic of the higher-energy LL switches from electron-like to hole-like LL while that of the lower-energy LL changes from hole-like to electron-like LL. As a result, the edge states change their behavior significantly across a hybridization gap.

Figure~\ref{fig:zeroFF}(d) illustrates edge state diagrams at $B_{c1}^<$ and $B_{c1}^>$, indicated by pink and orange lines in Fig.~\ref{fig:zeroFF}(c).  Due to a confining potential, the energy of an electron-like LL will bend up near the edge while that of a hole-like LL will bend down. At $B_{c1}^>$ [Fig.~$4$(d), right],  an energy ordering of the LLs is a conventional one in which the electron-like (B,~$1+$) has higher energy than the hole-like (b,~$1-$).  In this case, the  energies of both LLs will bend away from each other near the edge. Therefore, these two LLs do not contribute any edge state to the system at $\nu = 0$. However, at $B_{c1}^<$, we have an inverted energy ordering of the LLs in which the energy of hole-like (b,~$1-$) is higher than that of the electron-like (B,~$1+$) in the bulk. Near the edge, their energies will bend toward each other (down for (b,~$1-$)  and up for (B,~$1+$))  [Fig.~\ref{fig:zeroFF}(d), left].  We therefore obtain two counter-propagating edge states even though the filling factor is zero in the bulk. We note that these counter-propagating edge states are likely not helical since they can interact via mixing terms. As a result, we expect an energy gap to open at the crossing.

 As we lower $B$ further, we encounter another hybridization gap between  (b,~$2-$) and (B,~$2+$) at $B_{c2} \sim 1.2$~T [Fig.~$4$(a-c)]. With the same argument as the $B_{c1}$ case, the number of edge states will increase from $2$ to $4$ when $B$ drops below $B_{c2}$ because two hole-like (b,~$1-$) and (b,~$2-$) now sit at higher energy than two electron-like (B,~$1+$) and  (B,~$2+$). In general, as we move through the hybridization gap between  (b,~$i-$) and (B,~$i+$), the number of edge states changes from $2(i-1)$ to $2i$.  We find that the positions of the conductance peaks at $\nu = 0$ in Fig.~\ref{fig:zeroFF}(b) are in excellent agreement with the theoretical positions of $B_{ci}$ in Fig.~4(c) which are magnetic fields at which a slope is zero for a hybridized LL separating $\nu = 0$ and $4$. For $B > B_{c1}$, the energy gap at $\nu = 0$ turns into a trivial LL gap and there is no further inversion of electron-like and hole-like LLs for LL indices $|n| \geq 1$. Therefore, the number of edge states stays constant and we no longer observe any oscillation [Fig.~\ref{fig:zeroFF}(b)]. Similarly, following the $\nu = 12$ line in Fig.~\ref{fig:zeroFF}(a), we observe magnetoconductance oscillations when  $B  \lesssim 1.2$~T even though the $\nu = 12$ gap remains finite in the bulk. We find that this $\nu = 12$ gap below $1.2$~T arises from a series of hybridization gaps between (b,~$i-$) and (B,~$(i+3)+$) while the gap above $1.2$~T is a trivial LL gap.

We therefore conclude that the oscillations arise from the change in the number of edge states in a hybridization gap between electron-like and hole-like LLs. We emphasize that a hybridization gap between LLs of the same type will not lead to an oscillation because the number of edge states remains unchanged~\cite{Shimazaki2016LandauTrilayer,Zhang2006Anticrossing}. Our results show that it is not sufficient to predict Shubnikov-de Haas oscillations from Landau level spectra only. One needs to consider if a gap is a trivial Landau level gap or a hybridization gap between electron-like and hole-like LLs to obtain a complete picture of the oscillations. We expect our result to be useful for understanding Landau level spectra of other few layer graphene systems because of the band inversion in their band structures.

 For helical edge modes, two-terminal conductance in the hybridization gap regime along $\nu = 0$ should appear as steps with quantized values of $4N e^2/h$ where $N$ is the number of edge states and a factor of $4$ comes from spin and valley degeneracy~\cite{young2014tunable,sanchez2017helical,che2020helical}. For instance, when $\nu = 0$ and $B_{c2} < B < B_{c1}$, we expect conductance of $8e^2/h$ from two counter-propagating edge states from (b,~$1-$) and (B,~$1+$) but the measured value is less than $2 e^2/h$.  A few mechanisms could contribute to the observed low value of conductance. Since the edge states in our system are not helical and they counterpropagate on the same edge, these two states could mix and tunnel to each other.  As a result, they form 1D localized states and conductance is no longer quantized at $4N e^2/h$ because the edge states do not have a perfect transmission~\cite{Lee1985}.  An interaction between the edge states could induce a small gap at the Fermi energy, reducing conductance further. In addition, our conductance appears oscillatory which is likely due to the geometry effect observed in a long sample for two-terminal measurement~\cite{abanin2008conformal, williams2009quantum}.

In summary, we study magnetotransport properties of 4LG. We observe LL crossings between the b and B subbands. At finite displacement field, we find that the LLs in the b~subband become valley polarized while those in the B subband remain valley degenerate, in agreement with the TB calculation with finite potential difference. At low $n$ and $B$, the band inversion gives rise to a series of bulk hybridization gaps between electron- and hole-like LLs. As a result,  alternating characteristic of the hybridized LLs between electron and hole states leads to the change in the number of edge states and manifests as magnetoconductance oscillations in our measurement. Finally, our proposed mechanism for magnetoconductance oscillations should be applicable to other Bernal-stacked multilayer graphenes since they also host similar band inversion~\cite{Hirahara20184LGpeaks,Horii20194LG}.

\begin{acknowledgments}
We thank K. Jaruwongrungsee for experimental help and S. Hodak for useful discussion. This research has been primarily supported by the Research Fund for DPST graduate with First Placement (Grant no.~002/2015), the NSRF via the Program Management Unit for Human Resources \& Institutional Development, Research and Innovation (Grant no.~B05F640152), and National Research Council of Thailand (NRCT) and Chulalongkorn University (Grant no.~N42A650266).   K.W. and T.T. acknowledge support from JSPS KAKENHI (Grant Numbers 19H05790, 20H00354 and 21H05233).
\end{acknowledgments}

%
\newpage
\onecolumngrid
\renewcommand{\theequation}{S\arabic{equation}}
\renewcommand{\thefigure}{S\arabic{figure}}
\renewcommand{\thetable}{S\arabic{table}}
\renewcommand{\theenumi}{S\arabic{enumi}}
\renewcommand{\thesection}{S\Roman{section}}
\setcounter{figure}{0} 
\begin{center}
	\Large{\bf Supplemental Material}
\end{center}
{\bf Effect of Potential Difference on Landau Levels}

In our experiment, the potential difference between adjacent layers $U$ varies with carrier density but, in the calculation of Landau levels, we use a constant value of the potential difference. Therefore, the calculated Landau levels may not agree with the experimental data for a whole range of density values. One of the discrepancies is an absence of $\nu = 28$ gap at high field in our data. To account for it, we consider the effect of $U$ on the Landau level spectra.

Figures S1(a, b) and S1(c, d) show Landau level spectra for $U = 16$~meV and $U=22$~meV, respectively.    As $U$ becomes larger, the LL (b, 1+) shifts up in energy relatively to the LL (B, 5+). As a result,the energy at which LL (b, 1+) and LL (B, 5+) cross increases. With varying $U$, the LL (b, $1+$) could disperse along the LL (B, 5+) which leads to the absence of nu = 28 gap over a considerable range of magnetic field, similar to the data in Fig. 2(b).
\begin{figure}[H]
    \begin{center}
	\includegraphics[scale=1.2]{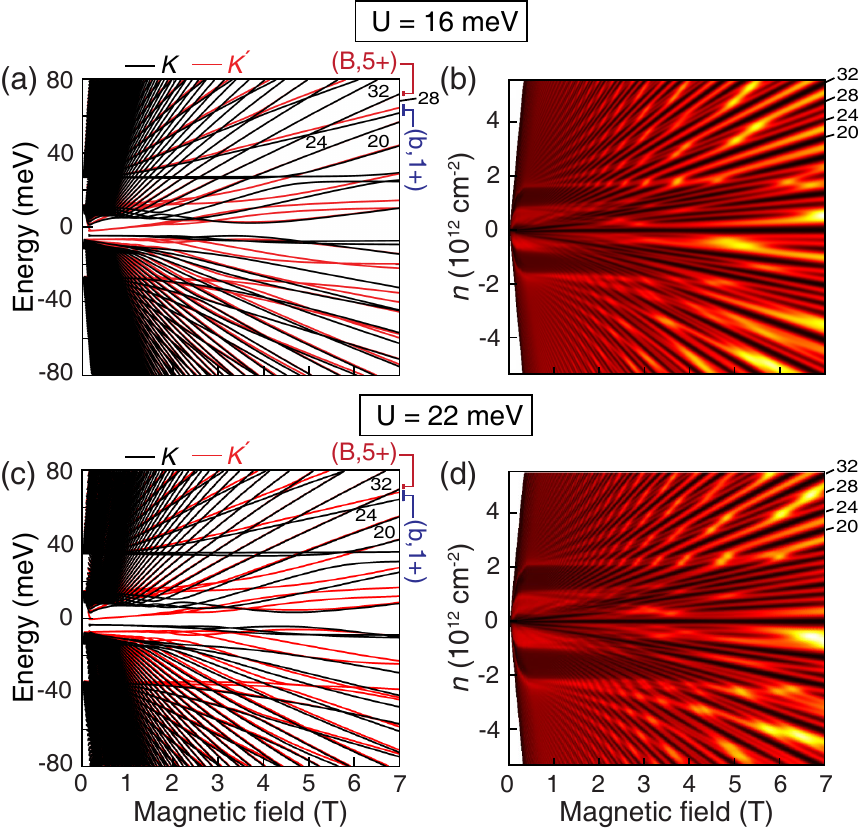}
\end{center}
	\caption{(a, b) Landau level spectra and the calculated density of states of the bulk 4LG as a function of n and B, using the potential difference between adjacent layers $U$ of $16$ meV. (c, d) Same with (a, b) but with $U = 22$~meV. The position of the crossing between (B, 5+) and (b, 1+) can be seen to shift up in energy which causes the absence of $\nu = 28$ gap at $7$~T.}
    
\end{figure}

\newpage

{\bf Additional Data on Unexpected Oscillations at Zero Density}

We have measured conductance on 3 devices [see Fig. S2], two of which are two-terminal devices (Device A and B) and the other is a four-terminal device (Device C). Figures 2 and 4 from the main text use data from Device A and Device B, respectively. All three devices display very similar oscillations in which positions of the peaks occur at almost identical values of magnetic field.

\begin{figure}[H]
    \begin{center}
	\includegraphics{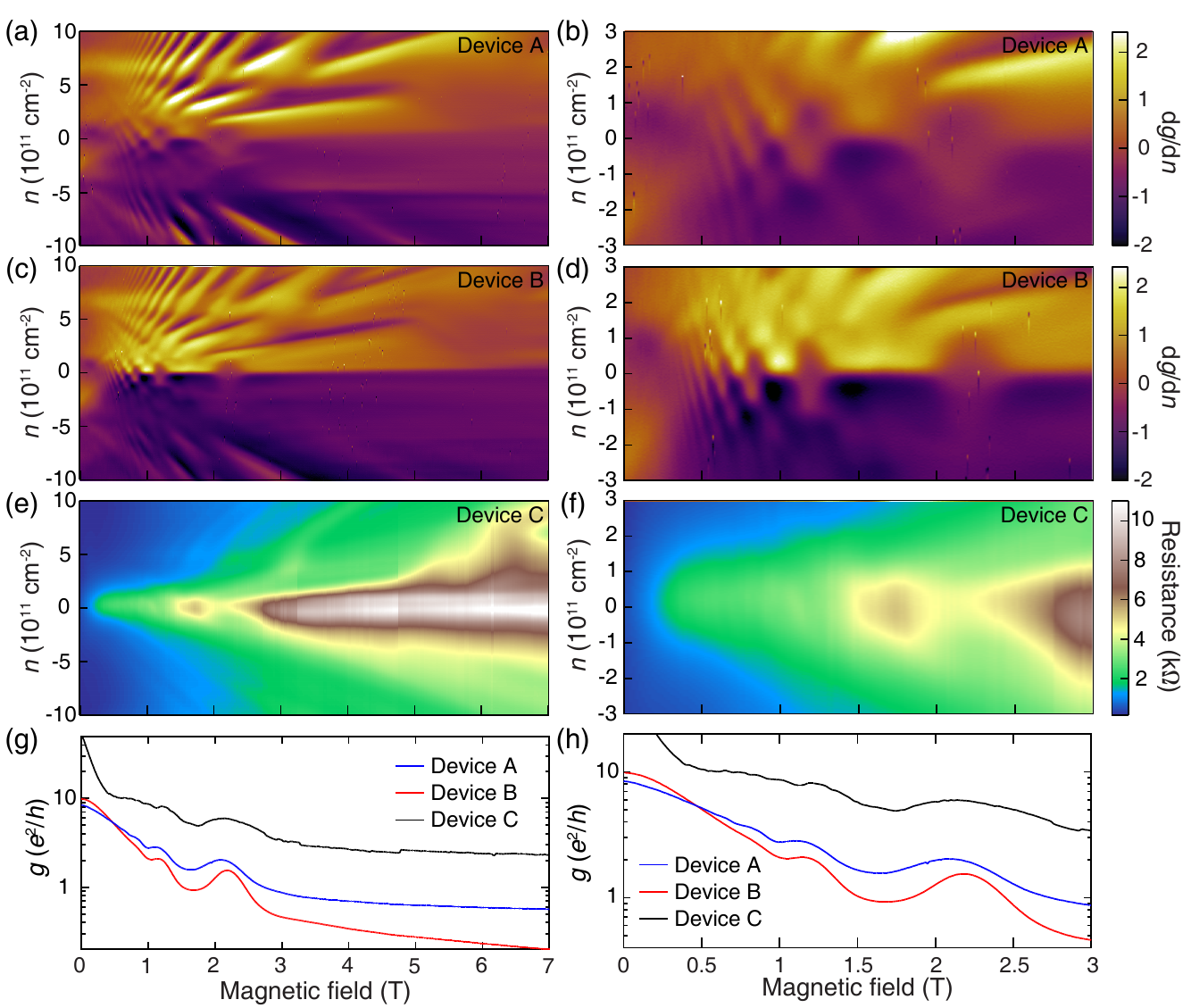}
    \end{center}
	\caption{(a, b) Color map of d$g$/d$n$ from device A at 0 -- 7 T and at 0 -- 3 T, respectively. (c, d) Color map of d$g$/d$n$ from device B at 0 -- 7 T and at 0 -- 3 T, respectively. (e, f) Color map of $R_{xx}$ measured in device C with 4-terminal geometry at 0 -- 7 T and at 0 -- 3 T, respectively. (g,~h) Conductance as a function of magnetic field at $n = 0$ from devices A (blue line), B (red line), and $1/R_{xx}$  as a function of magnetic field at $n = 0$ from device C (black line) at 0 -- 7~T and at 0~--~3~T, respectively. The oscillations in all 3 devices show consistent behaviors.}
\end{figure}

\newpage 

{\bf Periodicity in $1/B$ of magnetoconductance oscillations at $n=0$}

We observe that the magnetoconductance oscillations are periodic in inverse magnetic field, suggesting the origin related to Landau levels. We have included a plot of conductance oscillations with a smooth background subtracted as a function of inverse magnetic field (Fig.~S3).

\begin{figure}[H]
    \begin{center}
	\includegraphics{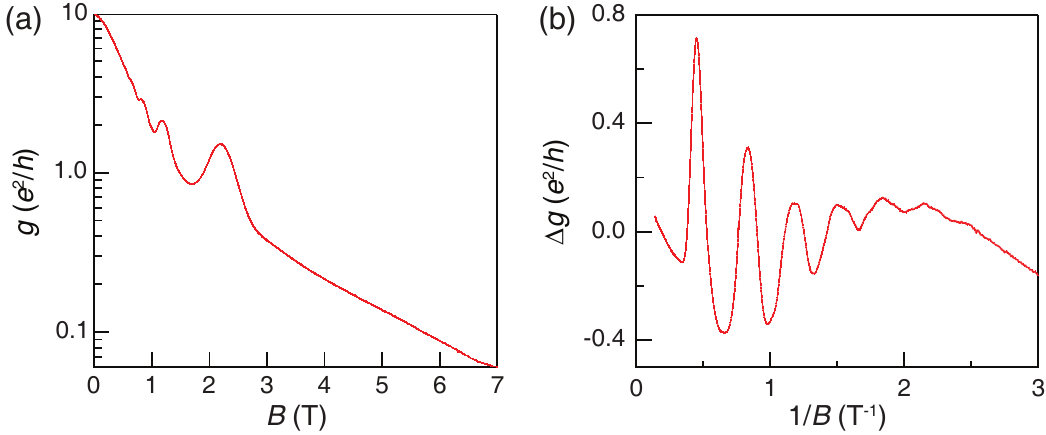}
    \end{center}
	\caption{(a) Conductance as a function of magnetic field at $n = 0$ and $T = 2.4$~K.  (b) Conductance from (a) with a smooth background subtracted as a function of inverse magnetic field.  }
\end{figure}

\end{document}